%% file: main.tex
\documentclass{arXiv_preprint}

\usepackage{graphicx}
\usepackage{wrapfig}
\usepackage{amsmath}
\usepackage{booktabs}
\usepackage{times}
\usepackage{amsfonts}
\usepackage{amssymb}
\usepackage{xcolor}
\usepackage{enumitem}
\usepackage{subcaption}
\usepackage{svg}
\usepackage{bm}
\usepackage{lipsum}
\usepackage{amsthm}
\usepackage{mathtools}
\usepackage{dsfont}
\usepackage{cancel}
\usepackage{algorithm}
\usepackage{algpseudocode}
\usepackage{float}
\usepackage{multirow}
\usepackage{array}
\usepackage{adjustbox}
\usepackage{accents}
\usepackage{cancel}
\usepackage{float}
\usepackage{booktabs}
\usepackage{longtable}
\usepackage{pdflscape}
\usepackage{makecell}
\usepackage{tikz}
\usepackage{minted}
\usepackage{dirtree}
\usepackage{fontawesome5}
\usepackage[bottom]{footmisc}
\usepackage{varwidth}
\usepackage{url}

\usepackage[
    backend=biber,
    style=numeric,
    sorting=none,
    url=false,
    doi=false,  
    isbn=false,
    eprint=false,
    giveninits=true,
    date=year
]{biblatex}
\AtEveryBibitem{\clearname{editor}}
\AtEveryBibitem{\clearlist{editor}}
\AtEveryBibitem{\clearlist{publisher}}
\AtEveryBibitem{\clearfield{publisher}}
\AtEveryBibitem{\clearlist{location}}
\AtEveryBibitem{\clearfield{location}}
\AtEveryBibitem{\clearfield{eventtitle}}
\AtEveryBibitem{\clearfield{venue}}
\AtEveryBibitem{\clearfield{series}}

\usepackage[
    hidelinks=true,
    colorlinks=true,
    linkcolor=black,
    urlcolor=black,
    citecolor=black
]{hyperref}
\usepackage[acronym,nonumberlist,nopostdot,nogroupskip,toc=true]{glossaries} 
\glsaddkey{value_range}{\glsentrytext{\glslabel}}{\glsentryvaluerange}{\GLsentryvaluerange}{\glsvaluerange}{\Glsvaluerane}{\GLSvaluerange}

\newglossarystyle{custom}{%
    \setglossarystyle{long3col} 

}

\glsdisablehyper

\DeclareMathOperator*{\argmin}{argmin}

\DeclarePairedDelimiter\abs{\lvert}{\rvert}

\algdef{SE}[DOWHILE]{Do}{doWhile}{\algorithmicdo}[1]{\algorithmicwhile\ #1}%

\definecolor{progressforeground}{RGB}{0,0,0}
\definecolor{progressbackground}{RGB}{255,255,255}

\newcommand{\progresscircle}[1]{%
    \begin{tikzpicture}[baseline=-0.65ex]
        \def\radius{0.15}
        \fill[progressbackground] (0,0) circle (\radius);
        \ifnum#1=100
            \fill[progressforeground] (0,0) circle (\radius);
        \else
            \ifnum#1>0
                \pgfmathsetmacro{\endangle}{90-3.6*(#1)}
                \fill[progressforeground]
                    (0,0) --
                    (90:\radius)
                    arc[
                        start angle=90,
                        end angle=\endangle,
                        radius=\radius
                    ]
                    -- cycle;
            \fi
        \fi
        \draw[black, line width=0.25pt] (0,0) circle (\radius);
    \end{tikzpicture}%
}

\definecolor{CodeBackground}{HTML}{F7F7F7}
\definecolor{CodeFrame}{HTML}{D9D9D9}

\definecolor{VSCodeKeyword}{HTML}{AF00DB}

\definecolor{VSCodeClass}{HTML}{267F99}

\definecolor{VSCodeFunction}{HTML}{795E26}

\definecolor{VSCodeVariable}{HTML}{001080}

\definecolor{VSCodeString}{HTML}{A31515}

\definecolor{VSCodeNumber}{HTML}{098658}

\definecolor{VSCodeComment}{HTML}{008000}

\definecolor{VSCodeConstant}{HTML}{0000FF}

\definecolor{VSCodeOperator}{HTML}{000000}

\newcommand{\PyKeyword}[1]{%
    \textcolor{VSCodeKeyword}{\ttfamily\detokenize{#1}}%
}

\newcommand{\PyClass}[1]{%
    \textcolor{VSCodeClass}{\ttfamily\detokenize{#1}}%
}

\newcommand{\PyFunction}[1]{%
    \textcolor{VSCodeFunction}{\ttfamily\detokenize{#1}}%
}

\newcommand{\PyVariable}[1]{%
    \textcolor{VSCodeVariable}{\ttfamily\detokenize{#1}}%
}

\newcommand{\PyString}[1]{%
    \textcolor{VSCodeString}{%
        \ttfamily
        \char`\"%
        \detokenize{#1}%
        \char`\"%
    }%
}

\newcommand{\PyNumber}[1]{%
    \textcolor{VSCodeNumber}{\ttfamily\detokenize{#1}}%
}

\newcommand{\PyComment}[1]{%
    \textcolor{VSCodeComment}{%
        \ttfamily
        \char`\#\hspace{0.6em}%
        \detokenize{#1}%
    }%
}

\newcommand{\PyConstant}[1]{%
    \textcolor{VSCodeConstant}{\ttfamily\detokenize{#1}}%
}

\setminted[python]{
    style=bw,
    bgcolor=CodeBackground,
    fontsize=\footnotesize,
    fontfamily=tt,
    breaklines=false,
    autogobble=true,
    tabsize=4,
    frame=single,
    framesep=0.5mm,
    rulecolor=CodeFrame,
    baselinestretch=1.05,
    linenos=false
}

\newsavebox{\PythonCodeBox}
\newsavebox{\PseudocodeBox}

\newcommand{\PseudocodeScale}{0.8}

\newcommand{\PseudocodeInternalScale}{%
    \fpeval{1 / (\PseudocodeScale)}%
} 
\newcommand{\PrintPseudocodeBox}[1]{%
    \par
    \vspace{0.125\baselineskip}
    \noindent
    \makebox[\linewidth][c]{%
        \scalebox{\PseudocodeScale}{#1}%
    }%
    \newline
    \par
}

\newcommand{\PseudocodeWidth}[1]{%
    \dimexpr\fpeval{#1/\PseudocodeScale}\linewidth\relax
}

\definecolor{foldercolor}{HTML}{D6A300}
\definecolor{filecolor}{HTML}{5B6770}
\definecolor{treecommentcolor}{HTML}{6A737D}

\newcommand{\foldericon}{%
    \textcolor{foldercolor}{\faFolderOpen}\hspace{0.4em}%
}

\newcommand{\codefileicon}{%
    \textcolor{filecolor}{\faFileCode}\hspace{0.4em}%
}

\newcommand{\tighttt}[2][0.70]{%
    \scalebox{#1}[1]{\texttt{#2}}%
}

\makeatletter
\def\namedlabel#1#2{\begingroup
    #2%
    \def\@currentlabel{#2}%
    \phantomsection\label{#1}\endgroup
}
\makeatother

\newcommand{\FootnoteUrl}[1]{%
    {\scriptsize\url{#1}}%
}

\makenoidxglossaries
\loadglsentries{glossary}

\title{\tighttt{D\MakeLowercase{istributed}D\MakeLowercase{esign}O\MakeLowercase{ptimizer}}: A modular Python framework for Setup, Execution and Processing of Distributed Design Optimization}
\vspace{4pt}

\author{Sebastian Ellmaier$^{*\dag}$, Albert Jan de Wit$^{\ddag}$, Akilesh Raveendran$^{\dag}$, Marc-Eric Vogt$^{\dag}$, Thomas Bäck$^{*}$ and Anna V. Kononova$^{*}$}
\shortauthors{S. Ellmaier, A. J. de Wit, A. Raveendran, M.-E. Vogt, T. Bäck and A. V. Kononova}

\address{%
$^{*}$ Leiden Institute of Advanced Computer Science (LIACS)
\\Leiden University
\\Leiden, 2300 RA, The Netherlands
\\\{m.r.ellmaier, a.kononova\}@liacs.leidenuniv.nl
\\www.liacs.leidenuniv.nl\\
[12pt]
$^{\ddag}$ Royal Netherlands Aerospace Centre (NLR)
\\Postbus 90502, 1006 BM, Amsterdam, The Netherlands
\\www.nlr.nl\\
[12pt]
$^{\dag}$ BMW Group
\\Munich, Petuelring 130, Germany
\\www.bmwgroup.com}


\summary{ 
\acrfull{mdo} enables the application of optimization algorithms to complex engineered systems, ranging from aerospace and automotive to robotics and microelectronics. Such multi-component systems typically consist of coupled subsystems, each characterized by its own design variables, constraints and objectives. Without adequate coordination of these couplings, subsystems may be optimal in isolation while the overall system remains suboptimal or even infeasible.

Distributed design optimization coordinates coupled subsystem optimization problems while allowing them to retain control over their local design variables. Although promising coordination methods exist, their application is hindered by the lack of a comprehensive software framework which supports intuitive problem definition, provides suitable distributed coordination algorithms, is modular and extensible, enables interactive (post-)processing, and supports the distributed computation of subsystem executions. 
This work derives requirements for such software, assesses existing frameworks against them, and introduces \tighttt{DistributedDesignOptimizer}, an open-source Python framework to fill this gap. 

}

\begin{document}

\maketitle

\section{Introduction}
\label{sec:introduction} 

Engineered systems consist of coupled components and disciplines, such as structures, aerodynamics, and propulsion, whose design is distributed among specialized teams~\cite{isermannDesignsSpecification2023,wangNetworkTarget2012a,tosseramsDistributedOptimization2008}.
A suitable design optimization method should therefore coordinate coupled subsystem optimization algorithms, preserving their autonomy over constraints, objectives, and design decisions.

Popular monolithic approaches (e.g., \acrfull{idf} and \acrfull{mdf}) distribute the analysis of a design choice among coupled disciplines, and are implemented in software such as OpenMDAO~\cite{grayOpenMDAOOpensource2019} or GEMSEO~\cite{gallardGEMSPython2018}.
However, the decision making (i.e. choosing the value of design variables) is centralized in a single optimizer~\cite{tosseramsDistributedOptimization2008}.
Distributed design optimization instead allows subsystems to also retain control over local design variables and coordinates their decisions toward overall feasibility and optimality~\cite{martinsMultidisciplinaryDesign2013a,tosseramsDistributedOptimization2008}.

The widespread application of promising coordination approaches remains difficult because available implementations are often restricted to a particular algorithm, problem class, or application domain. A general-purpose software that combines algorithm-independent problem definition, interchangeable coordination methods, extensibility, benchmarking, and (post-)processing for distributed engineering design optimization is missing.

This work addresses this gap by deriving software requirements, assessing existing frameworks against them, and introducing the framework \tighttt{DistributedDesignOptimizer}. It separates the problem definition from the coordination method and provides a common modular structure for defining, executing, extending, and analyzing distributed design optimization problems and algorithms. This work introduces the framework code structure and includes an exemplary problem definition, algorithm execution, result analysis, and extensions with novel methods.

\section{Distributed Design Optimization}
\label{sec:distributed_design_optimization}

A subsystem~$i$ comprises \textit{\glsdesc{x}}~${}^{i}\gls{x}$, \textit{\glsdesc{z}s}~${}^{i}_{j}\gls{z}$ with neighboring subsystem~$j\in{}^{i}\gls{N}$, and \textit{\glsdesc{r}}~${}^{i}\gls{r}$ entering its objective and constraints\footnote{This section summarizes relevant material from the authors' publication~\citefield{ellmaierDistirbutedDesignMultidisciplinary}{title}~\cite{ellmaierDistirbutedDesignMultidisciplinary}.}:

\begin{subequations}
\vspace{-0.75\baselineskip}
\begin{align}
    ^{i}\gls{x}^{*},{}\left\{^{i}_{j}\gls{z}^{*}\right\}_{j\in{}^{i}\gls{N}} := & \argmin_{\substack{^{i}\gls{x}\:\in{}^{i}\gls{X} \\ {}^{i}_{j}\gls{z}\:\in{}^{i}_{j}\gls{Z}\:\forall\:j\in{}^{i}\gls{N}}} \quad {}^{i}\gls{vf}\left(^{i}\gls{r}\right)
    \\
    & \text{s.t.} \quad \quad \:{}^{i}\gls{vleq}\left(^{i}\gls{r}\right) \leq 0\:,
    \\
    & \phantom{\text{s.t.}} \quad \quad \:{}^{i}\gls{veq}\left(^{i}\gls{r}\right) = 0\:,
    \\
    & \text{where} \quad  {}^{i}\gls{r}
    :=
    {}^{i}\gls{r}\!\left(
    {}^{i}\gls{x},
    \left\{{}^{i}_{j}\gls{z},{}^{j}_{i}\gls{h}\right\}_{j\in{}^{i}\gls{N}}
    \right)\:.
\end{align}
\label{eq:definition_optimization_problem_subsystem_i}
\vspace{-0.75\baselineskip}
\end{subequations}

Notably, ${}^{i}\gls{r}$ depends on $^{i}\gls{x}$, $^{i}_{j}\gls{z}$, and responses received from neighboring subsystems via \textit{\glsdesc{H}} $^{j}_{i}\gls{H}\left(^{j}\gls{r}\right) =: {}^{j}_{i}\gls{h}$.
The collection of these distributed, yet coupled design optimization problems $i\in\gls{M}$ leads to the \acrfull{aio} optimization problem, where the shared design variables~$^{i}_{j}\gls{z}$~and~$^{j}_{i}\gls{z}$ are condensed into a single design variable $\gls{z}_{\{i,j\}}$:

\begin{subequations}  
\vspace{-0.75\baselineskip}
\begin{align}
    \left\{^{i}\gls{x}^{*}\right\}_{i\in\gls{M}},{}\left\{\gls{z}_{\{i,j\}}^{*}\right\}_{\substack{j\in{}^{i}\gls{N}\\i\in\gls{M}}} := & \argmin_{\substack{^{i}\gls{x}\:\in{}^{i}\gls{X} \: \forall \: i\in\gls{M} \\ \gls{z}_{\{i,j\}}\:\in{}^{i}_{j}\gls{Z}\:\cap\:^{j}_{i}\gls{Z}\:\forall j\in{}^{i}\gls{N},{}i\in\gls{M} }} \quad \sum_{i\in\gls{M}}{}^{i}\gls{vf}\left(^{i}\gls{r}\right)
    \label{eq:aio_with_coupling_identidy_argmin}
    \\
    & \text{s.t.} \quad \quad \: {}^{i}\gls{vleq}\left(^{i}\gls{r}\right) \leq 0 \quad \forall \: i\in\gls{M}\:,
    \label{eq:aio_with_coupling_identidy_local_inequality_constraints}
    \\
    & \phantom{s.t.} \quad \quad \: {}^{i}\gls{veq}\left(^{i}\gls{r}\right) = 0 \quad \forall \: i\in\gls{M}\:,
    \label{eq:aio_with_coupling_identidy_local_equality_constraints}
    \\
    & \text{where} \quad {}^{i}\gls{r} := {}^{i}\gls{r}\left(^{i}\gls{x},{}\left\{\gls{z}_{\{i,j\}},{}^{j}_{i}\gls{h}\right\}_{j\in{}^{i}\gls{N}}\right)  \quad \forall \: i\in\gls{M}\:,
    \label{eq:aio_with_coupling_identidy_responses}
    \\
    & \phantom{where} \quad {}^{j}_{i}\gls{h} := {}^{j}_{i}\gls{H}\left(^{j}\gls{r}\right) \quad \forall\:j\in{}^{i}\gls{N},{}i\in\gls{M}\:.
    \label{eq:aio_with_coupling_identidy_mapping_identity}
\end{align}
\label{eq:aio_with_coupling_identidy}
\vspace{-0.75\baselineskip}
\end{subequations}

Many strategies apply a \textit{decomposition} reformulation to Equation~\eqref{eq:aio_with_coupling_identidy} by treating $^{j}_{i}\gls{h}$ and $^{i}_{j}\gls{z}$ as additional design variables of subsystem $i$ and introducing pairwise \textit{\glsdesc{c}}
\begin{equation}
    ^{i}_{j}\gls{c}:=
    \begin{bmatrix}
        ^{i}_{j}\gls{c}_{h} \\
        ^{i}_{j}\gls{c}_{z}
    \end{bmatrix}
    :=
    \begin{bmatrix}
        ^{i}_{j}\gls{H}\left(^{i}\gls{r}\right) - {}^{i}_{j}\gls{h} \\
        ^{i}_{j}\gls{z} - {}^{j}_{i}\gls{z}
    \end{bmatrix}
    = 0 \quad \forall\:j\in{}^{i}\gls{N},{}i\in\gls{M}\:,
    \label{eq:coupling_constraints_between_i_and_j}
\end{equation}
while omitting redundant coupling constraints between \glsdesc{z}s, i.e., only one of 
$^{i}_{j}\gls{c}_{z}=0$ and $^{j}_{i}\gls{c}_{z}=0$ is imposed~\cite{dewitUnifiedApproach2009a,tosseramsDistributedOptimization2008}. 
This leads to the \acrfull{aao} problem
\begin{subequations}
\vspace{-0.43\baselineskip}
\begin{align}
    \left\{^{i}\gls{x}^{*}\right\}_{i\in\gls{M}},{}\left\{^{i}_{j}\gls{z}^{*},{}^{j}_{i}\gls{h}^{*}\right\}_{\substack{j\in{}^{i}\gls{N}\\i\in\gls{M}}} := & \argmin_{\substack{^{i}\gls{x}\:\in{}^{i}\gls{X} \: \forall \: i\in\gls{M} \\ 
    ^{i}_{j}\gls{z}\:\in{}^{i}_{j}\gls{Z}\: \forall \:j\in{}^{i}\gls{N},{}i\in\gls{M} \\
    ^{j}_{i}\gls{h}\:\in{}^{j}_{i}\gls{H_set}\: \forall \:j\in{}^{i}\gls{N},{}i\in\gls{M}}} \quad \sum_{i\in\gls{M}}{}^{i}\gls{vf}\left(^{i}\gls{r}\right)
    \label{eq:aao_argmin}
    \\
    & \text{s.t.} \quad \quad  \: {}^{i}\gls{vleq}\left(^{i}\gls{r}\right) \leq 0 \quad \forall \: i\in\gls{M}\:,
    \label{eq:aao_local_inequality_constraints}
    \\
    & \phantom{s.t.} \quad \quad  \: {}^{i}\gls{veq}\left(^{i}\gls{r}\right) = 0 \quad \forall \: i\in\gls{M}\:,
    \label{eq:aao_local_equality_constraints}
    \\
    & \phantom{s.t.} \quad \quad  \: {}^{i}_{j}\gls{c} = 0 \quad \forall \:j\in{}^{i}\gls{N},{}i\in\gls{M}\:,
    \label{eq:aao_coupling_constraints}
    \\
    & \text{where} \quad {}^{i}\gls{r} := {}^{i}\gls{r}\left(^{i}\gls{x},\left\{^{i}_{j}\gls{z},{}^{j}_{i}\gls{h}\right\}_{j\in{}^{i}\gls{N}}\right)  \quad \forall \: i\in\gls{M}\:.
    \label{eq:aao_responses}
\end{align}
\label{eq:aao}
\vspace{-0.75\baselineskip}
\end{subequations}

The optimal solutions of Equations~\eqref{eq:aio_with_coupling_identidy} and \eqref{eq:aao} coincide when the set $^{j}_{i}\gls{H_set}$ is selected appropriately.
In contrast to the \acrshort{aio} formulation of Equation~\eqref{eq:aio_with_coupling_identidy}, the above reformulation renders response $^{i}\gls{r}$ dependent only on \glsdesc{d}s $\left[ ^{i}\gls{x},\left\{^{i}_{j}\gls{z},{}^{j}_{i}\gls{h}\right\}_{j\in{}^{i}\gls{N}} \right] =: {}^{i}\gls{d}$ of subsystem~$i$.

From the perspective of multidisciplinary engineering organizations, a suitable coordination algorithm should satisfy three requirements:

\begin{description}[font=\normalfont\itshape, labelsep=0pt, itemsep=\parskip, parsep=0pt]
    \item[Req. \namedlabel{itm:req_no1}{No.1}]\textit{: Numerical Efficiency}
    \quad
    The algorithm yields a solution of high quality (i.e., optimality and feasibility) at reasonable computational expense.
    \item[Req. \namedlabel{itm:req_no2}{No.2}]\textit{: Wide Applicability}
    \quad
    The algorithm ideally covers general constrained nonlinear problems and settings involving limited derivative information.
    \item[Req. \namedlabel{itm:req_no3}{No.3}]\textit{: Distributed Coordination}
    \quad
    The algorithm allows subsystems to have autonomy over the definition of local constraints and objective as well as the choice of optimal local design variables, while couplings are handled systematically.    
\end{description}

The \acrshort{mdo} community developed various \textit{monolithic} or \text{distributed} solution algorithms~\cite{martinsMultidisciplinaryDesign2013a,tosseramsDistributedOptimization2008}.
Monolithic approaches such as \acrshort{idf} and \acrshort{mdf} permit distributed analyses but retain centralized optimization, thereby only partially representing the autonomy of distributed engineering organizations. 
Distributed approaches instead assign coordinated optimization problems to the subsystems and better support organizationally distributed design.
Among distributed approaches, an additional distinction concerns whether subsystem autonomy over local and shared variables is preserved and whether bidirectional coupling structures can be handled.
\acrshort{bliss} (2000) assigns shared decisions to a controller, while \acrshort{csso} modifies subsystem problems by controller-defined factors. Thus, both of these algorithms are unsuitable for the considered setting as they impair subsystem autonomy. \acrshort{co} and \acrshort{atc} are limited to unidirectional couplings. 
Consequently, the most suitable classes for the considered distributed design optimization setting are relaxation-based primal-dual methods (such as \acrfull{alc}~\cite{tosseramsDistributedOptimization2008}, \acrshort{consensus_alc}~\cite{wangNetworkTarget2012a}, and \acrshort{aladin}~\cite{houskaAugmentedLagrangian2016}), and \acrfull{sbdp}~\cite{voneschSensitivityBasedDistributed2025}.

Despite their methodological differences, these approaches share a common algorithmic principle.
Each subsystem repeatedly solves an optimization problem consisting of its local objective and constraints, augmented by an additional \textit{\glsdesc{P}} $^{i}\gls{P}$ and/or \textit{\glsdesc{Q}} $^{i}\gls{Q}$ parametrized by \textit{\glsdesc{u}} \gls{u}, which depends on information received from neighboring subsystems and, where applicable, from a dedicated controller entity.
After each solve, selected coupling quantities are exchanged and updated until convergence.
This observation motivates the unified algorithmic structure of Alg.~\ref{algo:unified_algorithmic_structure_part_2}, which includes relevant distributed primal-dual and sensitivity-based methods as specializations.

\begin{algorithm}
\caption{Unified Algorithmic Structure} \label{algo:unified_algorithmic_structure_part_2}
\begin{lrbox}{\PseudocodeBox} 
\begin{minipage}{\PseudocodeInternalScale\linewidth}
\begin{algorithmic}[1]    
    \Require hyperparameters for inner- and outer-loop convergence criteria
    \Statex \hspace{0.975cm} hyperparameters for update of relevant coupling parameters
    \Statex \hspace{0.975cm} initial relevant coupling parameters in interface 
    \State $\gls{outer-loop_itr} \leftarrow 0$  \Comment{initialize \glsdesc{outer-loop_itr}}
    \Do
        \State $\gls{inner-loop_itr} \leftarrow 0$ \Comment{initialize \glsdesc{inner-loop_itr}}
        \Do \Comment{following some iteration scheme}
            \For{\textbf{every subsystem} $i\in\gls{M}$}
                \State Copy relevant coupling parameters of $^{C}_{i}\gls{u},{}^{j}_{i}\gls{u}\:\forall\: j \in{}^{i}\gls{N}$ from interface
                \State Prepare optimization problem formulation
                \State
                \vspace{-2.0em}
                \begin{equation}
                \begin{split}
                    ^{i}\gls{d}^{(\gls{outer-loop_itr},\gls{inner-loop_itr}+1)} \leftarrow & \argmin_{^{i}\gls{d}\:\in{}^{i}\gls{D_set}} \quad {}^{i}\gls{vf}\left(^{i}\gls{r}\right) + {}^{i}\gls{P}\left(^{i}\gls{d},{}\left\{^{i}_{j}\gls{u},{}^{j}_{i}\gls{u}\right\}_{j\in{}^{i}\gls{N}},{}^{i}_{C}\gls{u},{}^{C}_{i}\gls{u}\right)
                    \\
                    & \text{s.t.} \quad {}^{i}\gls{vleq}\left(^{i}\gls{r}\right) \leq 0\:,
                    \\
                    & \phantom{\text{s.t.} \quad} {}^{i}\gls{veq}\left(^{i}\gls{r}\right) = 0\:,
                    \\
                    & \phantom{\text{s.t.} \quad} {}^{i}\gls{Q}^{\leq}\left(^{i}\gls{d},{}\left\{^{i}_{j}\gls{u},{}^{j}_{i}\gls{u}\right\}_{j\in{}^{i}\gls{N}},{}^{i}_{C}\gls{u},{}^{C}_{i}\gls{u}\right) \leq 0\:,
                    \\
                    & \phantom{\text{s.t.} \quad} {}^{i}\gls{Q}^{=}\left(^{i}\gls{d},{}\left\{^{i}_{j}\gls{u},{}^{j}_{i}\gls{u}\right\}_{j\in{}^{i}\gls{N}},{}^{i}_{C}\gls{u},{}^{C}_{i}\gls{u}\right) = 0\:.
                \end{split}
                \label{eq:unified_structure_optimization_problem}
                \end{equation}
                \vspace{-1.0em}
                \State Post-process the optimization
                \State Copy relevant coupling parameters of $^{i}_{C}\gls{u},{}^{i}_{j}\gls{u}\:\forall\: j \in{}^{i}\gls{N}$ to interface
            \EndFor

\algstore{myalg}

\end{algorithmic}
\end{minipage} 
\end{lrbox}
\noindent \scalebox{\PseudocodeScale}{%
    \usebox{\PseudocodeBox}%
}
\end{algorithm}

\begin{algorithm}
\begin{lrbox}{\PseudocodeBox} 
\begin{minipage}{\PseudocodeInternalScale\linewidth}
\begin{algorithmic} [1]
\algrestore{myalg}
            
            \For{\textbf{Controller}}
                \State Copy relevant coupling parameters of $^{i}_{C}\gls{u}\: \forall\: i \in \gls{M}$ from interface
                \State Prepare optimization problem formulation
                \State
                \vspace{-2.0em}
                \begin{equation}
                \begin{split}
                    ^{C}\gls{d} \leftarrow & \argmin_{^{C}\gls{d}\:\in{}^{C}\gls{D_set}\left(\left\{^{C}_{i}\gls{u},{}^{i}_{C}\gls{u}\right\}_{i\in\gls{M}}\right)} \quad  ^{C}\gls{P}\left(^{C}\gls{d},{}\left\{^{C}_{i}\gls{u},{}^{i}_{C}\gls{u}\right\}_{i\in\gls{M}}\right)
                    \\
                    & \text{s.t.} \quad {}^{C}\gls{Q}^{\leq}\left(^{C}\gls{d},{}\left\{^{C}_{i}\gls{u},{}^{i}_{C}\gls{u}\right\}_{i\in\gls{M}}\right) \leq 0\:,
                    \\
                    & \phantom{\text{s.t.} \quad} {}^{C}\gls{Q}^{=}\left(^{C}\gls{d},{}\left\{^{C}_{i}\gls{u},{}^{i}_{C}\gls{u}\right\}_{i\in\gls{M}}\right) = 0\:.
                \end{split}
                \label{eq:unified_structure_controller_optimization_problem}
                \end{equation}
                \vspace{-1.0em}
                \State Post-process the optimization
                \State Copy relevant coupling parameters of $^{C}_{i}\gls{u}\: \forall\: i \in \gls{M}$ to interface
            \EndFor
            \For{\textbf{every subsystem} $i\in\gls{M}$ \textbf{and Controller in parallel}}
                \State Copy relevant coupling parameters of $^{C}_{i}\gls{u},{}^{j}_{i}\gls{u}\:\forall\: j \in{}^{i}\gls{N}$ and $^{i}_{C}\gls{u}\:\forall\:i\in\gls{M}$ from interface,~resp.
                \State Update relevant coupling parameters of $^{i}_{C}\gls{u},{}^{C}_{i}\gls{u},{}^{i}_{j}\gls{u},{}^{j}_{i}\gls{u}\:\forall\: j\in{} ^{i}\gls{N}$ and $^{i}_{C}\gls{u},{}^{C}_{i}\gls{u}\:\forall\:i\in\gls{M}$,~resp.
                \State Copy relevant coupling parameters of $^{i}_{C}\gls{u},{}^{i}_{j}\gls{u}\:\forall\: j \in {}^{i}\gls{N}$ and $^{C}_{i}\gls{u}\:\forall\:i\in\gls{M}$ to interface,~resp. 
            \EndFor
            \State Compute inner-loop convergence criterion
            \State $\gls{inner-loop_itr} \leftarrow \gls{inner-loop_itr}+1$
        \doWhile{inner-loop convergence criterion is False}
                
        \Statex
        \For{\textbf{every subsystem} $i\in\gls{M}$ \textbf{and Controller in parallel}}
            \State Copy relevant coupling parameters of $^{C}_{i}\gls{u},{}^{j}_{i}\gls{u}\:\forall\: j\in{}^{i}\gls{N}$ and $^{i}_{C}\gls{u}\:\forall\:i\in\gls{M}$ from interface,~resp.
            \State Prepare update of relevant coupling parameters of $^{i}_{C}\gls{u},{}^{C}_{i}\gls{u},{}^{i}_{j}\gls{u},{}^{j}_{i}\gls{u}\:\forall\: j \in{}^{i}\gls{N}$ and $^{i}_{C}\gls{u},{}^{C}_{i}\gls{u}\:\forall\:i\in\gls{M}$,~resp.
            \State Copy relevant coupling parameters of $^{i}_{C}\gls{u},{}^{i}_{j}\gls{u}\:\forall\: j \in{}^{i}\gls{N}$ and $^{C}_{i}\gls{u}\:\forall\:i\in\gls{M}$ to interface,~resp.
        \EndFor
        \For{\textbf{every subsystem} $i\in\gls{M}$ \textbf{and Controller in parallel}}
            \State Copy relevant coupling parameters of $^{C}_{i}\gls{u},{}^{j}_{i}\gls{u}\:\forall\: j\in{}^{i}\gls{N}$ and $^{i}_{C}\gls{u}\:\forall\:i\in\gls{M}$ from interface,~resp.
            \State Update relevant coupling parameters of $^{i}_{C}\gls{u},{}^{C}_{i}\gls{u},{}^{i}_{j}\gls{u},{}^{j}_{i}\gls{u}\:\forall\: j \in{}^{i}\gls{N}$ and $^{i}_{C}\gls{u},{}^{C}_{i}\gls{u}\:\forall\:i\in\gls{M}$,~resp.
            \State Copy relevant coupling parameters of $^{i}_{C}\gls{u},{}^{i}_{j}\gls{u}\:\forall\: j \in{}^{i}\gls{N}$ and $^{C}_{i}\gls{u}\:\forall\:i\in\gls{M}$ to interface,~resp.
        \EndFor
        \State Compute outer-loop convergence criterion
        \State $\gls{outer-loop_itr} \leftarrow \gls{outer-loop_itr} + 1$
    \doWhile{outer-loop convergence criterion is False}
    \State \Return $\left\{^{i}\gls{d}^{(\gls{outer-loop_itr})}\right\}_{i\in\gls{M}}$
\end{algorithmic}
\end{minipage} 
\end{lrbox}
\noindent \scalebox{\PseudocodeScale}{%
    \usebox{\PseudocodeBox}%
}
\end{algorithm}

\section{Lack of Suitable Comprehensive Software}
\label{sec:open_issues}

Although distributed primal-dual optimization and \acrshort{sbdp} are deemed most promising, they are not widely used in engineering organizations~\cite{jelevPatternSearch2019a}\footnote{This section summarizes relevant material from the authors' publication~\citefield{ellmaierDistirbutedDesignMultidisciplinary}{title}~\cite{ellmaierDistirbutedDesignMultidisciplinary}.}.
As emphasized in~\cite{papageorgiouRoleMultidisciplinary2017a,tosseramsDistributedOptimization2008}, adoption also requires usability, transparency, and integration into engineering workflows. The lack of software support for defining, executing, and analyzing distributed design optimization is a major open issue.
A suitable framework should offer the following six features~\cite{schwarzPycity_schedulingAPython2021,boydDistributedOptimization2011,dewitUnifiedApproach2009a,huangExtensibleMultiagent2006a}:

\begin{description}[font=\normalfont\itshape, labelsep=0pt, itemsep=\parskip, parsep=0pt]
    \item[Feature \namedlabel{itm:feature_no1}{No.1}]\textit{: Problem Definition}
    \quad 
    Design engineers can formulate their distributed design optimization problem independent of the chosen algorithm.
    
    \item[Feature \namedlabel{itm:feature_no2}{No.2}]\textit{: Choice of Algorithm}
    \quad
    Design engineers can choose and execute any suitable distributed optimization algorithm mentioned in Sec.~\ref{sec:distributed_design_optimization}.
    
    \item[Feature \namedlabel{itm:feature_no3}{No.3}]\textit{: Modular \& Extensible Framework}
    \quad 
    Algorithm developers can easily implement novel distributed optimization algorithms within a modular software framework.
    
    \item[Feature \namedlabel{itm:feature_no4}{No.4}]\textit{: Suite of Standardized Benchmark Problems}
    \quad 
    Design engineers and algorithm developers can test and compare algorithms on a standardized set of benchmark problems.
    
    \item[Feature \namedlabel{itm:feature_no5}{No.5}]\textit{: (Post-)Processing \& Analysis Capabilities}
    \quad
    Design engineers and algorithm developers can process and analyze the algorithm during execution and gain insights.
    
    \item[Feature \namedlabel{itm:feature_no6}{No.6}]\textit{: Local \& Cluster Deployment}
    \quad
    The chosen distributed optimization algorithm is deployable on a local machine or a distributed computation cluster, e.g., via \acrshort{mpi}.
\end{description}

Existing software only partially satisfies these requirements. Research code released to reproduce publications is typically problem-specific and unsuitable for productive use. The generic \acrshort{admm} code by~\citeauthor{boydDistributedOptimization2011}~\cite{boydDistributedOptimization2011} and the modular
\textit{ADMM framework of Sutor}~\cite{sutorAlternatingDirection2015} provide implementation insights, but they are limited to linear coupling constraints.

More mature frameworks are often domain-specific: \textit{GRAMPC-D} addresses distributed optimal control~\cite{burkModularFramework2022}, \textit{ProxAL} targets AC optimal power flow~\cite{subramanyamGloballyConvergent2021}, and \textit{PycityScheduling} focuses on energy scheduling~\cite{schwarzPycity_schedulingAPython2021}.
Although they provide modular setup, execution, and post-processing, they do not address distributed design optimization of this work.
The \textit{ProximalAlgorithms.jl} package provides modular implementations of several related distributed optimization algorithms~\cite{themelisDouglasRachford2022}. However, it is restricted to problems with two or three additive objectives.

The \textit{pyMDO} framework supports structured problem definition and generates executable reformulations for monolithic and distributed approaches, including \acrshort{co}, \acrshort{csso}, and \acrshort{bliss}~\cite{martinsPyMDOObjectOriented2009}.
Its object-oriented code is extensible, and its logging captures convergence histories and runtimes. Nevertheless, it lacks the methods and unified algorithmic structure considered here, does not support parallel processing, and is no longer publicly available.
\textit{OpenMDAO} and \textit{GEMSEO} offer comprehensive functionality for defining, solving, and post-processing \acrshort{mdo} problems, but primarily target monolithic optimization approaches~\cite{grayOpenMDAOOpensource2019,gallardGEMSPython2018}. \textit{GEMSEO} supports a selection of distributed approaches~\cite{gazaixIndustrializationNew2017,davidBilevelArchitectures2024},
but these are incompatible with the requirements of Sec.~\ref{sec:distributed_design_optimization}.

Several frameworks are closer to the present work but remain limited. \textit{ALADIN-$\alpha$} supports problem setup, parallel subsystem execution, and online processing for \acrshort{aladin}-type methods, but is largely script-based and restricted to linear coupling constraints~\cite{engelmannALADINAnOpensource2022}.
\textit{DISROPT} offers a modular, object-oriented architecture, distributed execution, and several algorithms~\cite{farinaDISROPTPython2020}. However, its setup and analysis capabilities are limited, and its structure differs from Alg.~\ref{algo:unified_algorithmic_structure_part_2}. 
Dedicated environments such as
\textit{atcPortal}/\textit{atcEngine}~\cite{huangExtensibleMultiagent2006a}
and the \textit{ALC Matlab toolbox}~\cite{tosseramsUsingALC2009a}
address relevant aspects of distributed
design optimization but are either no longer publicly available or
difficult to extend beyond a single coordination method.
Related implementations \textit{NoHiMDO}\footnote{\FootnoteUrl{https://github.com/bastientalgorn/NoHiMDO}, \quad \FootnoteUrl{https://github.com/khbalhandawi/NoHiMDO}}, \textit{DMDO}\footnote{\FootnoteUrl{https://github.com/Ahmed-Bayoumy/DMDO}}, and
\textit{NHATC}\footnote{\FootnoteUrl{https://github.com/johnmartins/nhatc}} support distributed problem definition, online processing, and parallel execution but are largely tailored to
\acrshort{alc}~\cite{talgornCompactImplementation2017a}. The \textit{mlprogram} framework of~\cite{dewitUnifiedApproach2009a}
separates subsystem models, coordination mechanisms, message passing, and post-processing similarly to the unified structure derived in this work. Its problem
definition is independent of the solution approach and it
supports distributed execution; however, only \acrshort{alc} is fully  implemented.

In summary, existing frameworks provide valuable building blocks, but none sufficiently covers all Feat. \ref{itm:feature_no1}-\ref{itm:feature_no6} (Tab.~\ref{tab:distributed_optimization_frameworks}).
This hinders the productive use of distributed optimization in engineering organizations~\cite{jelevPatternSearch2019a,papageorgiouRoleMultidisciplinary2017a,bilMultidisciplinaryDesign2015a}. To fill this gap, a novel framework is introduced.

\begin{table}[H]
\vspace{-0.5\baselineskip}
\caption[Assessment of distributed optimization frameworks]%
{Assessment of distributed optimization frameworks
against the identified features:
\protect\progresscircle{100}~fully,
\protect\progresscircle{50}~partially, and
\protect\progresscircle{0}~not satisfied.
Frameworks incompatible with the considered distributed design optimization
formulation or algorithms are~marked~by~*.}
\centering

\scalebox{\PseudocodeScale}{%

\setlength{\tabcolsep}{3pt}

\begin{tabular}{
    >{\raggedright\arraybackslash}m{6.5cm}
    *{6}{>{\centering\arraybackslash}m{1.0cm}}
}
\toprule
\textbf{Framework \quad \quad \quad \quad \quad \quad \:\: Feature}
& \textbf{\ref{itm:feature_no1}}
& \textbf{\ref{itm:feature_no2}}
& \textbf{\ref{itm:feature_no3}}
& \textbf{\ref{itm:feature_no4}}
& \textbf{\ref{itm:feature_no5}}
& \textbf{\ref{itm:feature_no6}}
\\
\midrule

Boyd ADMM~\cite{boydDistributedOptimization2011}
& \progresscircle{25}
& \progresscircle{25}
& \progresscircle{25}
& \progresscircle{0}
& \progresscircle{25}
& \progresscircle{50}
\\

Sutor ADMM~\cite{sutorAlternatingDirection2015}
& \progresscircle{25}
& \progresscircle{25}
& \progresscircle{50}
& \progresscircle{0}
& \progresscircle{50}
& \progresscircle{25}
\\

\addlinespace[7pt]

GRAMPC-D~\cite{burkModularFramework2022}
& \progresscircle{0}*
& \progresscircle{25}
& \progresscircle{50}
& \progresscircle{25}
& \progresscircle{50}
& \progresscircle{50}
\\

ProxAL~\cite{subramanyamGloballyConvergent2021}
& \progresscircle{0}*
& \progresscircle{25}
& \progresscircle{50}
& \progresscircle{25}
& \progresscircle{25}
& \progresscircle{75}
\\

PycityScheduling~\cite{schwarzPycity_schedulingAPython2021}
& \progresscircle{0}*
& \progresscircle{50}
& \progresscircle{75}
& \progresscircle{50}
& \progresscircle{50}
& \progresscircle{75}
\\

ProximalAlgorithms.jl~\cite{themelisDouglasRachford2022}
& \progresscircle{0}*
& \progresscircle{50}
& \progresscircle{75}
& \progresscircle{25}
& \progresscircle{50}
& \progresscircle{50}
\\

\addlinespace[7pt]

pyMDO~\cite{martinsPyMDOObjectOriented2009}
& \progresscircle{25}
& \progresscircle{0}*
& \progresscircle{75}
& \progresscircle{50}
& \progresscircle{50}
& \progresscircle{25}
\\

OpenMDAO~\cite{grayOpenMDAOOpensource2019},
GEMSEO~\cite{gallardGEMSPython2018}
& \progresscircle{75}
& \progresscircle{0}*
& \progresscircle{75}
& \progresscircle{50}
& \progresscircle{75}
& \progresscircle{75}
\\

\addlinespace[7pt]

ALADIN-$\alpha$~\cite{engelmannALADINAnOpensource2022}
& \progresscircle{25}
& \progresscircle{50}
& \progresscircle{25}
& \progresscircle{25}
& \progresscircle{50}
& \progresscircle{50}
\\

DISROPT~\cite{farinaDISROPTPython2020}
& \progresscircle{50}
& \progresscircle{50}
& \progresscircle{75}
& \progresscircle{25}
& \progresscircle{50}
& \progresscircle{50}
\\

atcPortal / atcEngine~\cite{huangExtensibleMultiagent2006a}
& \progresscircle{75}
& \progresscircle{25}
& \progresscircle{50}
& \progresscircle{25}
& \progresscircle{75}
& \progresscircle{75}
\\

ALC Matlab toolbox~\cite{tosseramsUsingALC2009a}
& \progresscircle{50}
& \progresscircle{25}
& \progresscircle{25}
& \progresscircle{25}
& \progresscircle{25}
& \progresscircle{25}
\\

NoHiMDO, DMDO, NHATC
& \progresscircle{50}
& \progresscircle{25}
& \progresscircle{50}
& \progresscircle{50}
& \progresscircle{50}
& \progresscircle{50}
\\

mlprogram~\cite{dewitUnifiedApproach2009a}
& \progresscircle{50}
& \progresscircle{25}
& \progresscircle{75}
& \progresscircle{50}
& \progresscircle{50}
& \progresscircle{50}
\\

\addlinespace[7pt]

\textbf{DistributedDesignOptimizer}
& \progresscircle{50}
& \progresscircle{75}
& \progresscircle{87}
& \progresscircle{50}
& \progresscircle{87}
& \progresscircle{50}
\\

\bottomrule
\end{tabular}

}

\medskip

\label{tab:distributed_optimization_frameworks}
\vspace{-0.5\baselineskip}
\end{table}

\section{\tighttt{D\MakeLowercase{istributed}D\MakeLowercase{esign}O\MakeLowercase{ptimizer}} Framework}
\label{sec:the_framework}

The \tighttt{DistributedDesignOptimizer} is a modular Python framework for the setup, execution, and processing of distributed design
optimization. It is designed to offer Feats.~\ref{itm:feature_no1}-\ref{itm:feature_no6} of Section~\ref{sec:open_issues}, and is published under the GNU LGPLv3 License\footnote{\FootnoteUrl{https://www.gnu.org/licenses/lgpl-3.0.html}} on GitHub\footnote{\FootnoteUrl{https://github.com/SebastianEllmaier/DistributedDesignOptimizer}}. Its documentation supplements the following sections with methodological and architectural details and tutorials.

\subsection{Architecture Overview} \label{sec:architecture-overview}

\begin{wrapfigure}{r}{0.405\textwidth}
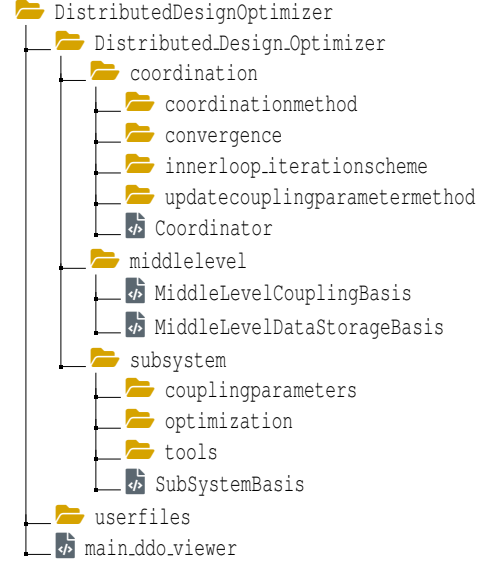

    \centering
    \begin{lrbox}{\PseudocodeBox} 
        \begin{minipage}{\PseudocodeInternalScale\linewidth} 
            \dirtree{%
                .1 \foldericon \tighttt{DistributedDesignOptimizer}. 
                .2 \foldericon \tighttt{Distributed\_Design\_Optimizer}.
                .3 \foldericon \tighttt{coordination}.
                .4 \foldericon \tighttt{coordinationmethod}.
                .4 \foldericon \tighttt{convergence}.
                .4 \foldericon \tighttt{innerloop\_iterationscheme}.
                .4 \foldericon \tighttt{updatecouplingparametermethod}.
                .4 \codefileicon \tighttt{Coordinator}.
                .3 \foldericon \tighttt{middlelevel}.
                .4 \codefileicon \tighttt{MiddleLevelCouplingBasis}.
                .4 \codefileicon \tighttt{MiddleLevelDataStorageBasis}.
                .3 \foldericon \tighttt{subsystem}.
                .4 \foldericon \tighttt{couplingparameters}.
                .4 \foldericon \tighttt{optimization}.
                .4 \foldericon \tighttt{tools}.
                .4 \codefileicon \tighttt{SubSystemBasis}.
                .2 \foldericon \tighttt{userfiles}.
                .2 \codefileicon \tighttt{main\_ddo\_viewer}.
            }
        \end{minipage}
    \end{lrbox}
    \scalebox{\PseudocodeScale}{%
        \usebox{\PseudocodeBox}%
    }
    \caption{Directory structure of the 
    \tighttt{DistributedDesignOptimizer}.}
    \label{fig:repository-directory-tree}
    \vspace{-0.5\baselineskip}
\end{wrapfigure}

The fundamental packages, classes, and scripts of the framework are shown in Fig.~\ref{fig:repository-directory-tree}.
To address Feat.~\ref{itm:feature_no3}, the modular implementation of
algorithms is realized through \tighttt{coordination}, \tighttt{subsystem}, and \tighttt{middlelevel}. Benchmark problems are included in \tighttt{userfiles} for Feat.~\ref{itm:feature_no4}.
Logging data is stored in \tighttt{userfiles/<problem>/historyfiles} and processed using \tighttt{main\_ddo\_viewer}.

The \tighttt{coordination} package contains the \tighttt{Coordinator}, which initializes the algorithm, spawns outer- and inner-loops of lines 2, 4, 26, and 40 in Alg.~\ref{algo:unified_algorithmic_structure_part_2}, and invokes subsystem calls of lines 5, 12, 19, 28, and 33 in Alg.~\ref{algo:unified_algorithmic_structure_part_2}.
The  \tighttt{coordinationmethod} package holds classes for algorithm initial-
ization.
The \tighttt{convergence}, \tighttt{innerloop\_iterationsscheme}, and \tighttt{updatecouplingparametermethod} packages provide interchangeable classes for different convergence criteria of lines 24 and 38, iteration schemes, e.g., sequential or parallel execution via \tighttt{multiprocessing}, and update strategies of lines 21 and 35 in Alg.~\ref{algo:unified_algorithmic_structure_part_2}.

The \tighttt{subsystem} package represents the individual processing units of
the distributed algorithm. Common subsystem behavior is implemented in \tighttt{SubSystemBasis}, from which coordination-method-specific classes are
derived. Each \text{subsystem} maintains its \text{optimization} problem of Equation~\eqref{eq:unified_structure_optimization_problem}, coupling
parameters, and numerical tools.
The \tighttt{couplingparameters} package holds method-specific classes for coupling parameter~$\gls{u}$.
The \tighttt{optimization} package provides analysis, optimization problem, and
solver abstractions to handle the subsystem optimization. Local terms ${}^{i}\gls{r}$, ${}^{i}\gls{vf}$, ${}^{i}\gls{vleq}$ are defined by \tighttt{Analysis<id>}, \tighttt{LocalObjective<id>}, and \tighttt{LocalConstraint<id>} classes in \tighttt{userfiles}. 
\tighttt{Analysis<id>} also maps responses and design variables to the coupling parameters. Although \tighttt{Analysis<id>} follows a strict template, it permits general Python scripts to evaluate ${}^{i}\gls{r}({}^{i}\gls{d})$.
$^{i}\gls{P}$ and $^{i}\gls{Q}$ are implemented by coordination-method-specific child classes of \tighttt{SubSystemBasis}.
Each subsystem may use a different numerical solver in \tighttt{Optimization<id>} in \tighttt{userfiles} to solve Equation~\eqref{eq:unified_structure_optimization_problem}.
Some third-party solvers are included, and alternatives can be added.
A coordination method may use constraint multipliers even when the selected
solver does not return them. In this case, \tighttt{SubSystemBasis} reconstructs them from the first-order \acrshort{kkt} stationarity conditions of Equation~\eqref{eq:unified_structure_optimization_problem}. 
Supporting functionality such as finite-difference Jacobian computation and \acrshort{bfgs} Hessian approximation is provided by \tighttt{tools}.

\begin{figure}[H]
    \vspace{-0.5\baselineskip}
    \centering
    \includegraphics[trim={0.7cm 0.5cm 0.7cm 9.4cm},clip,width=1.0\linewidth]{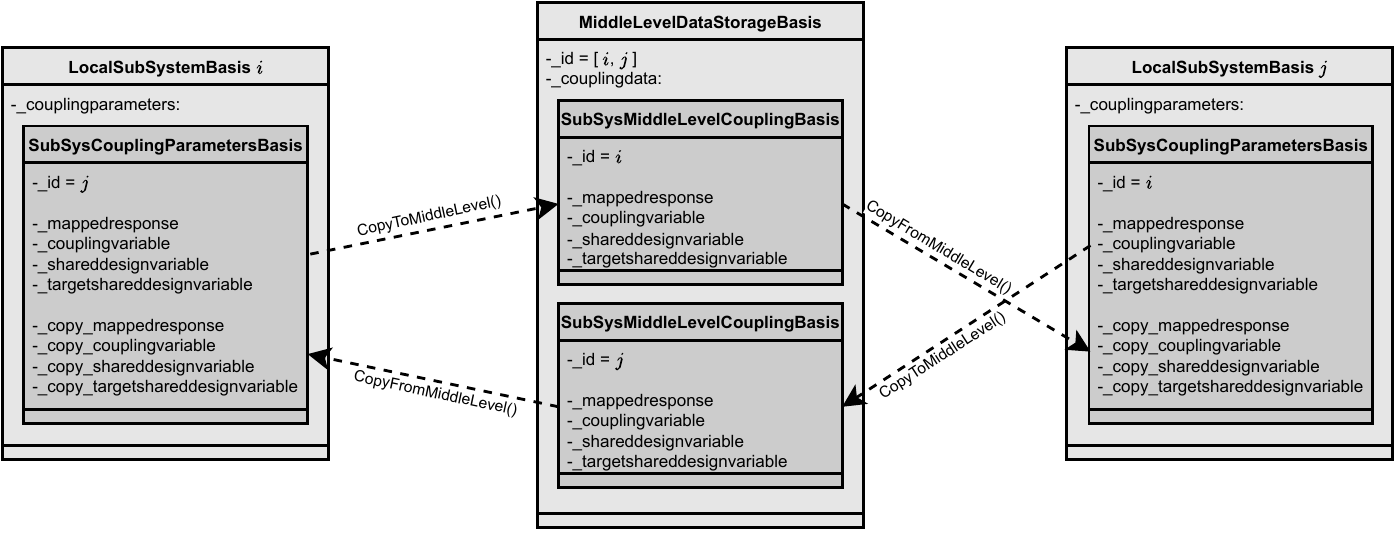}
    \caption{Information exchange between two \tighttt{LocalSubSystemBasis} instances via a \tighttt{MiddleLevelDataStorageBasis} instance.}
    \label{fig:MiddleLevelDataStorage_Two_LocalSubSystemBasis}
    \vspace{-0.5\baselineskip}
\end{figure}

While a subsystem's \glsdesc{u} $\gls{u}$ is represented by classes in \tighttt{couplingparameters}, the corresponding shared resource is provided by the \tighttt{middlelevel} package. Its \linebreak \tighttt{MiddleLevelDataStorageBasis} is the communication interface storage between neighboring \linebreak \tighttt{LocalSubSystemBasis} instances, each of which keeps a  \tighttt{SubSysCouplingParameterBasis} instance per pairwise coupling and copies values to/from the shared \tighttt{SubSysMiddleLevelCouplingBasis} of \linebreak \tighttt{MiddleLevelDataStorageBasis}. To address Feat.~\ref{itm:feature_no6}, access to each storage is protected by a \tighttt{multiprocessing} lock.
Coordination methods within the unified algorithmic structure seek to satisfy the coupling constraints $^{i}_{j}\gls{c}$ of Equation~\eqref{eq:coupling_constraints_between_i_and_j} between two \tighttt{LocalSubSystemBasis} instances.
Thus, the \tighttt{SubSysMiddleLevelCouplingBasis} between them holds attributes \tighttt{\_mappedresponses}~${}^{i}_{j}\gls{H}$, \tighttt{\_couplingvariable}~${}^{i}_{j}\gls{h}$, \tighttt{\_shareddesignvariable}~${}^{i}_{j}\gls{z}$, and \tighttt{\_targetshareddesignvariable}~${}^{i}_{j}\gls{z}_{t}$ as shown in Fig.~\ref{fig:MiddleLevelDataStorage_Two_LocalSubSystemBasis}.
Additional coordination-method-specific quantities may be exchanged. The middlelevel instance may also hold coupling data of two different types for asymmetric exchange.

\subsection{Problem Definition} \label{sec:setup_and_execution}

\begin{figure}[H]
    \vspace{-0.5\baselineskip}
    \centering
    \includegraphics[trim={0.7cm 0.7cm 0.7cm 0.7cm},clip,width=1.0\linewidth]{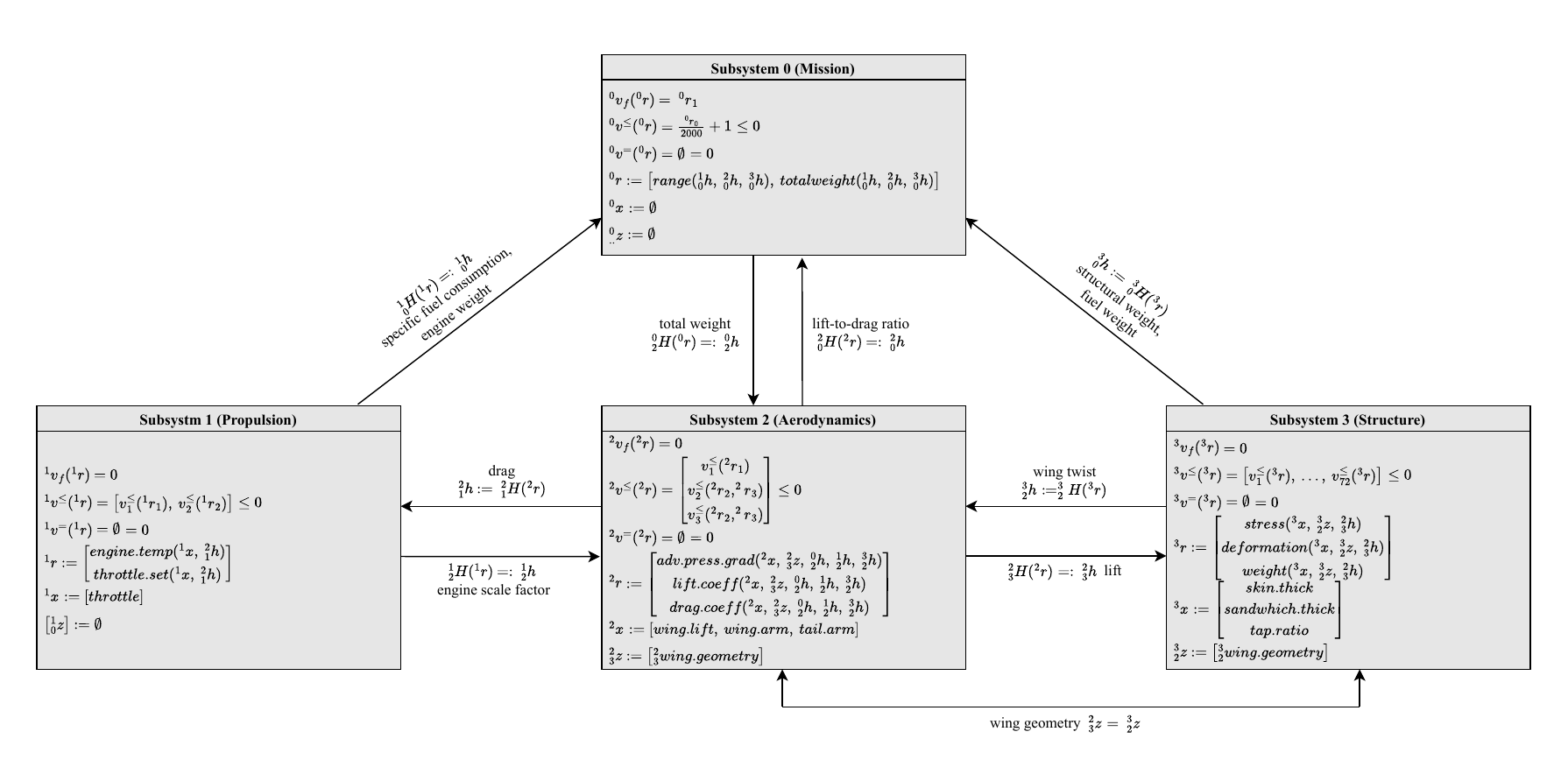}
    \caption{Distributed design optimization problem of the \acrshort{ssbj} design~\cite{talgornNumericalInvestigation2017a}.    
    }
    \label{fig:FIGURE_SSBJ_distributed_optimization_problem_modelling}
    \vspace{-0.5\baselineskip}
\end{figure}

As required by Feat.~\ref{itm:feature_no1}, the optimization problem definition is independent of the coordination method implementations. A \acrfull{ssbj} problem (adapted from~\cite{talgornNumericalInvestigation2017a},
shown in Fig.~\ref{fig:FIGURE_SSBJ_distributed_optimization_problem_modelling}) exemplifies the setup.

Each problem is implemented in a dedicated subdirectory of
\tighttt{userfiles} containing an \tighttt{InputFile},
an executable \tighttt{main}, and one directory per subsystem. The
subsystem directories hold implementations of the analysis,
objective, constraints, and local optimizer. 
The \tighttt{InputFile} defines all information necessary to set up and solve a distributed design optimization problem. Representative excerpts are shown for the \acrshort{ssbj} problem.
To address Feat.~\ref{itm:feature_no2}, the framework currently offers \acrshort{alc}, \acrshort{consensus_alc}, \acrshort{aladin}, and \acrshort{sbdp}. The selected class is instantiated with interchangeable convergence criteria, iteration schemes, and update strategies:

\begin{lrbox}{\PythonCodeBox} 
\begin{minipage}{\PseudocodeWidth{0.89}}
\begin{minted}[
    style=bw,
    escapeinside=||
]{python}
|\PyKeyword{class}| |\PyClass{InputFile}|(|\PyClass{InputFileBasis}|):
    |\PyKeyword{def}| |\PyFunction{__init__}|(|\PyVariable{self}|) -> |\PyConstant{None}|:
        |\PyFunction{super}|().|\PyFunction{__init__}|()
        |\PyComment{Name used for logging and post-processing}|
        |\PyVariable{self._name}| = |\PyString{SSBJ}|
        |\PyVariable{self._coordinationmethod}| = |\PyClass{ALC}|(convergence_indicator_inner-loop=...,
                                       convergence_indicator_outer-loop=...,
                                       updatecouplingparametermethod_outer-loop=...,
                                       iterationscheme=|\PyClass{SequentialForward}|())
\end{minted}
\end{minipage}
\end{lrbox}
\PrintPseudocodeBox{\usebox{\PythonCodeBox}}

Each subsystem is assigned a unique identifier $i$ and its
neighbors' identifiers $^{i}\gls{N}$. 
The imported \tighttt{Analysis<id>}, \tighttt{LocalObjective<id>}, \tighttt{LocalConstraint<id>}, and \tighttt{Optimization<id>} classes are instantiated. 
Design space ${}^{i}\gls{D_set}$ is defined by lower and upper bounds.
To account for different magnitudes among quantities, scaling similar to \textit{OpenMDAO}\footnote{\FootnoteUrl{https://openmdao.org/newdocs/versions/latest/features/core_features/working_with_components/scaling.html}} is used.
The granularities and initial values of $^{i}\gls{d}$ are defined next. 
For the aircraft subsystem of the \acrshort{ssbj} problem, this reads:

\begin{lrbox}{\PythonCodeBox} 
\begin{minipage}{\PseudocodeWidth{0.89}}
\begin{minted}[
    style=bw,
    escapeinside=||
]{python}
|\PyVariable{self._subsystems}|[|\PyNumber{0}|].|\PyFunction{set_DesignVariables_Granularity}|([|\PyNumber{0.0}|] * |\PyNumber{5}|)  |\PyComment{all continuous}|
|\PyVariable{self._subsystems}|[|\PyNumber{0}|].|\PyFunction{set_DesignVariables}|([
                        |\PyVariable{scalers}|[|\PyNumber{0}|].|\PyFunction{transform}|(|\PyNumber{2.0}|),       |\PyComment{fuel consumption [1/hr]}|
                        |\PyVariable{scalers}|[|\PyNumber{1}|].|\PyFunction{transform}|(|\PyNumber{15000.0}|),   |\PyComment{engine weight [lb]}|
                        |\PyVariable{scalers}|[|\PyNumber{2}|].|\PyFunction{transform}|(|\PyNumber{5.0}|),       |\PyComment{lift-to-drag ratio [-]}|
                        |\PyVariable{scalers}|[|\PyNumber{3}|].|\PyFunction{transform}|(|\PyNumber{25000.0}|),   |\PyComment{structural weight [lb]}|
                        |\PyVariable{scalers}|[|\PyNumber{4}|].|\PyFunction{transform}|(|\PyNumber{25000.0}|)])  |\PyComment{fuel weight [lb]}|
\end{minted}
\end{minipage}
\end{lrbox}
\PrintPseudocodeBox{\usebox{\PythonCodeBox}}

Neighbors may be coupled by \glsdesc{z} $^{i}_{j}\gls{z}$ and \glsdesc{h} \text{$^{j}_{i}\gls{h} := {}^{j}_{i}\gls{H}\left({}^{j}\gls{r}\right)$}. To define the coupling, the attributes of \tighttt{SubSysCouplingParameterBasis}  (Fig.~\ref{fig:MiddleLevelDataStorage_Two_LocalSubSystemBasis}) are initialized.
For example, fuel consumption and engine weight are \glsdesc{h}s $^{1}_{0}\gls{h}$ of the aircraft subsystem and mapped responses $^{1}_{0}\gls{H}\left(^{1}\gls{r}\right)$ of the propulsion
subsystem in the \acrshort{ssbj} problem:
\begin{lrbox}{\PythonCodeBox} 
\begin{minipage}{\PseudocodeWidth{0.89}}
\begin{minted}[
    style=bw,
    escapeinside=||
]{python}
|\PyComment{Aircraft-side representation of the coupling 0 -> 1}|
|\PyVariable{self._subsystems}|[|\PyNumber{0}|].|\PyFunction{set_CouplingVariables}|(
                    id=|\PyString{1}|,  |\PyComment{Propulsion}|
                    couplingvariablein=[
                        |\PyVariable{self._subsystems}|[|\PyNumber{0}|].|\PyFunction{get_Scalers}|()[|\PyNumber{0}|].|\PyFunction{transform}|(|\PyNumber{2.0}|),
                        |\PyVariable{self._subsystems}|[|\PyNumber{0}|].|\PyFunction{get_Scalers}|()[|\PyNumber{1}|].|\PyFunction{transform}|(|\PyNumber{15000.0}|)],
                    couplingvariablein_unscaled=[|\PyNumber{2.0}|,     |\PyComment{fuel consumption [1/hr]}|
                                                 |\PyNumber{15000.0}|  |\PyComment{engine weight [lb]}|])
|\PyVariable{self._subsystems}|[|\PyNumber{0}|].|\PyFunction{set_Copy_MappedResponseVariables}|(
                    id=|\PyString{1}|,  |\PyComment{Propulsion}|
                    copymappedresponsesin=[
                        |\PyVariable{self._subsystems}|[|\PyNumber{1}|].|\PyFunction{get_Scalers}|()[|\PyNumber{6}|].|\PyFunction{transform}|(|\PyNumber{2.0}|),
                        |\PyVariable{self._subsystems}|[|\PyNumber{1}|].|\PyFunction{get_Scalers}|()[|\PyNumber{7}|].|\PyFunction{transform}|(|\PyNumber{15000.0}|)])
\end{minted}
\end{minipage}
\end{lrbox}
\PrintPseudocodeBox{\usebox{\PythonCodeBox}}
The reciprocal coupling from the propulsion subsystem perspective is defined analogously.

\subsection{Processing} \label{sec:processing}

To address Feat.~\ref{itm:feature_no5}, the framework records algorithmically required information and data for analyzing algorithm behavior. 
Subsystem and system-wide information is stored in the \tighttt{subsystemhistory} and \tighttt{coordinatorhistory} attributes of \tighttt{SubSystemBasis} and \tighttt{Coordinator}.
Both are \tighttt{deque} objects to which snapshots are appended in each inner-loop iteration of Alg.~\ref{algo:unified_algorithmic_structure_part_2}.
Each snapshot is a \tighttt{HistoryEntry} containing loop indices, runtime, number of design variable evaluations, optimal design variables, coupling parameters, and convergence flags.

\begin{figure}[H]
\vspace{-0.5\baselineskip}
    \centering
    \includegraphics[scale=0.31]{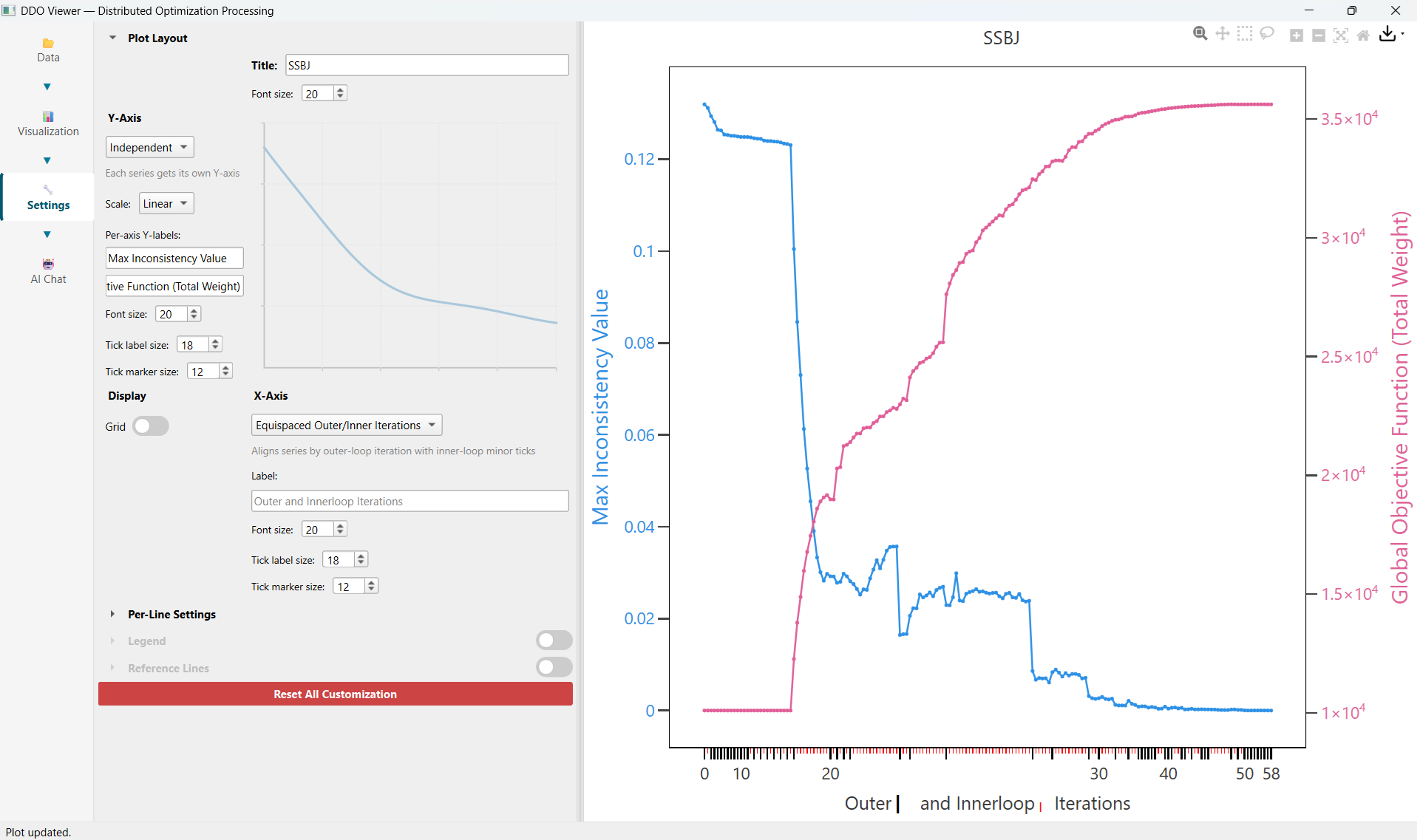}
    \caption{Screenshot of the \tighttt{DDO Viewer} settings panel and line plots for the \acrshort{ssbj} problem.}
    \label{fig:Screenshot_DDO_Viewer_Visualization_Panel_Line_Plot}
    \vspace{-0.5\baselineskip}
\end{figure}

The histories are accessible during execution and periodically serialized as \tighttt{.dill} files in \tighttt{userfiles/<problem>/historyfiles} for processing with \tighttt{main\_ddo\_viewer}. 
The \tighttt{DDO Viewer} enables visualization, analysis, and comparison of data from one or multiple executions.
Selected \tighttt{.dill} files are loaded into a data panel and refreshed during execution. A data-structure window lists plottable quantities for selection.
The settings panel offers customizations for further insights as illustrated in Fig.~\ref{fig:Screenshot_DDO_Viewer_Visualization_Panel_Line_Plot}, which plots the maximum coupling constraint violation (inconsistency) $\max_{i\in\gls{M},{}j\in{}^{i}\gls{N}}\abs{^{i}_{j}\gls{c}}$ and the global objective function value $\sum_{i\in\gls{M}}{}^{i}\gls{vf}$ for the \acrshort{ssbj} problem. 

This evolution is typical of relaxation-based coordination algorithms such as \acrshort{alc}:
Initially, coupling constraint violations are weakly penalized, allowing subsystems to optimize virtually in isolation. As the coordination drives the subsystems to consistency, the objective increases toward the best known optimum of $3.36\cdot{}10^{4}$. Hyperparameter tuning can further improve convergence~\cite{talgornNumericalInvestigation2017a}.

\subsection{Extending with a Novel Algorithm} \label{sec:extenting_novel_algorithm}

\begin{wrapfigure}[18]{r}{0.58\textwidth}
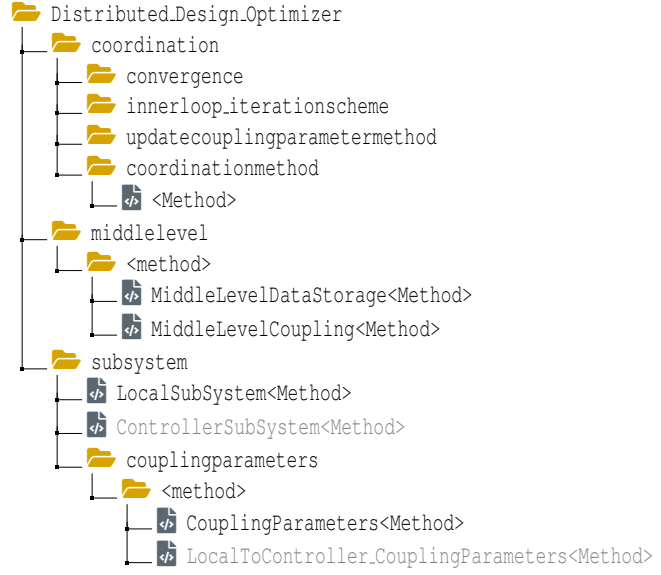

    \vspace{-\baselineskip}
    \centering
    \begin{lrbox}{\PseudocodeBox} 
        \begin{minipage}{\PseudocodeInternalScale\linewidth} 
            \dirtree{%
                .1 \foldericon \tighttt{Distributed\_Design\_Optimizer}.
                .2 \foldericon \tighttt{coordination}.
                .3 \foldericon \tighttt{convergence}.
                .3 \foldericon \tighttt{innerloop\_iterationscheme}.
                .3 \foldericon \tighttt{updatecouplingparametermethod}.
                .3 \foldericon \tighttt{coordinationmethod}.
                .4 \codefileicon \tighttt{<Method>}.
                .2 \foldericon \tighttt{middlelevel}.
                .3 \foldericon \tighttt{<method>}.
                .4 \codefileicon \tighttt{MiddleLevelDataStorage<Method>}.
                .4 \codefileicon \tighttt{MiddleLevelCoupling<Method>}.
                .2 \foldericon \tighttt{subsystem}.
                .3 \codefileicon \tighttt{LocalSubSystem<Method>}.
                .3 \codefileicon \textcolor{gray}{\tighttt{ControllerSubSystem<Method>}}.
                .3 \foldericon \tighttt{couplingparameters}.
                .4 \foldericon \tighttt{<method>}.
                .5 \codefileicon \tighttt{CouplingParameters<Method>}.
                .5 \codefileicon \textcolor{gray}{\tighttt{LocalToController\_CouplingParameters<Method>}}.            
            }
            \end{minipage}
    \end{lrbox}

    \scalebox{\PseudocodeScale}{%
        \usebox{\PseudocodeBox}%
    }
    
    \caption{Classes of a novel distributed optimization method. \textcolor{gray}{Gray} classes are required only for controllers.}
    \label{fig:novel-method-components}
    \vspace{-0.5\baselineskip}
\end{wrapfigure}
As required by Feat.~\ref{itm:feature_no3}, novel distributed optimization methods following Alg.~\ref{algo:unified_algorithmic_structure_part_2} can be added.
Common base classes execute the main algorithmic steps and provide fundamental functionality as detailed in Sec.~\ref{sec:architecture-overview}. 
If a novel \tighttt{<Method>} requires a new convergence criterion, iteration scheme, or parameter update strategy, the packages \tighttt{convergence}, \tighttt{innerloop\_iterationscheme}, and \linebreak \tighttt{updatecouplingparametermethod} need to be extended. 
If the \tighttt{<Method>} further utilizes a novel coupling parameter $\gls{u}$ and coordination terms $^{i}\gls{P}$ and $^{i}\gls{Q}$, the method-specific classes shown in Fig.~\ref{fig:novel-method-components} need to be implemented.

\section{Conclusion}
\label{sec:conclusion}

Distributed design optimization coordinates coupled subsystems toward system feasibility and optimality.
Its adoption requires software supporting modular implementation of novel methods, benchmark-based evaluation, structured problem definition, distributed execution, and comprehensive \text{(post-)processing}.
Since existing frameworks lack some identified features, \linebreak \tighttt{DistributedDesignOptimizer} is introduced for distributed primal-dual and sensitivity-based coordination methods. 
Its architecture separates subsystem optimization, coordination, and information exchange while allowing design problems to be defined independently of the coordination method. 
It also records optimization histories and provides interactive \text{(post-)processing}. 

The framework is under development to improve the features listed in Tab.~\ref{tab:distributed_optimization_frameworks}. 
Priorities include dedicated variable classes, a stronger separation of coordination, recording, and processing, and deployment on computational clusters. 
The benchmark suite should be expanded to improve algorithm validation and comparison.
Furthermore, the \tighttt{DDO Viewer} should become a holistic \acrshort{gui} for problem definition, algorithm configuration, execution, and processing.

\tighttt{DistributedDesignOptimizer} is envisioned as a standard open-source framework for solving practical problems and for implementing and comparing coordination methods.
Community contributions to its benchmarks, methodological capabilities, and usability are appreciated.

\section*{Acknowledgement}
We gratefully acknowledge Albert Jan de Wit for generously sharing his \textit{mlprogram}~\cite{dewitUnifiedApproach2009a}, which laid the foundation for the \tighttt{DistributedDesignOptimizer} framework. 



\printbibliography

\end{document}

%% file: glossary.tex






\newacronym{admm}{ADMM}{Alternating Directions Method of Multipliers}
\newacronym{ssbj}{SSBJ}{Super Sonic Business Jet}
\newacronym{FEM}{FEM}{Finite Element Method}
\newacronym{CFD}{CFD}{Computational Fluid Dynamics}
\newacronym{mdo}{MDO}{Multidisciplinary Design Optimization}
\newacronym{aao}{AAO}{All-At-Once}
\newacronym{aio}{AIO}{All-In-One}
\newacronym{idf}{IDF}{Individual Feasible}
\newacronym{mdf}{MDF}{Multidisciplinary Feasible}
\newacronym{fpi}{FPI}{Fixed Point Iteration}
\newacronym{co}{CO}{Collaborative Optimization}
\newacronym{csso}{CSSO}{Concurrent SubSpace Optimization}
\newacronym{bliss}{BLISS}{Bi-Level Integrated System Synthesis}
\newacronym{bliss2000}{BLISS-2000}{Bi-Level Integrated System Synthesis 2000}
\newacronym{atc}{ATC}{Analytical Target Cascading}
\newacronym{btc}{BTC}{Bounded Target Cascading}
\newacronym{alc}{ALC}{Augmented Lagrangian Coordination}
\newacronym{consensus_alc}{Consensus ALC}{Consensus Augmented Lagrangian Coordination}
\newacronym{ald}{ALD}{Augmented Lagrangian Duality}
\newacronym{mm}{MM}{Method of Multipliers}
\newacronym{aladin}{ALADIN}{Augmented Lagrangian based Alternating Direction Inexact Newton}
\newacronym{consensus_aladin}{Consensus ALADIN}{Consensus Augmented Lagrangian based Alternating Direction Inexact Newton}
\newacronym{kkt}{KKT}{Karush-Kuhn-Tucker}
\newacronym{sqp}{SQP}{Sequential Quadratic Program}
\newacronym{sbdp}{SBDP}{Sensitivity-based Distributed Programming}
\newacronym{consensus_sbdp}{Consensus SBDO}{Consensus Sensitivity-based Distributed Programming}
\newacronym{mpi}{MPI}{Message Passing Interface}
\newacronym{qp}{QP}{Quadratic Program}
\newacronym{gui}{GUI}{Graphical User Interface}
\newacronym{nomad}{NOMAD}{Nonlinear Optimization with Mesh Adaptive Direct Search}
\newacronym{bfgs}{BFGS}{Broyden–Fletcher–Goldfarb–Shanno}

\newglossaryentry{x}{
    name={\ensuremath{x}},
    description={local design variable},
    value_range={$\mathbb{R}^{\gls{nx}}$},
    sort={x}}
\newglossaryentry{nx}{
    name={\ensuremath{n_{\gls{x}}}},
    description={dimension of~\glsdesc{x}},
    value_range={$\mathbb{I}^{+}$},
    sort={nx}}
\newglossaryentry{X}{
    name={\ensuremath{\mathcal{X}}},
    description={set of~\glsdesc{x}},
    value_range={$\mathbb{R}^{\gls{nx}}$},
    sort={X}}
\newglossaryentry{z}{
    name={\ensuremath{z}},
    description={shared design variable},
    value_range={$\mathbb{R}^{\gls{nz}}$},
    sort={z}}
\newglossaryentry{nz}{
    name={\ensuremath{n_{\gls{z}}}},
    description={dimension of~\glsdesc{z}},
    value_range={$\mathbb{I}^{+}$},
    sort={nz}}
\newglossaryentry{Z}{
    name={\ensuremath{\mathcal{Z}}},
    description={set of~\glsdesc{z}},
    value_range={$\mathbb{R}^{\gls{nz}}$},
    sort={Z}}
\newglossaryentry{h}{
    name={\ensuremath{h}},
    description={coupling variable},
    value_range={$\mathbb{R}^{\gls{nh}}$},
    sort={h}}
\newglossaryentry{H_set}{
    name={\ensuremath{\mathcal{H}}},
    description={set of~\glsdesc{h}},
    value_range={$\mathbb{R}^{\gls{nh}}$},
    sort={H}}
\newglossaryentry{nh}{
    name={\ensuremath{n_{\gls{h}}}},
    description={dimension of~\glsdesc{h}},
    value_range={$\mathbb{I}^{+}$},
    sort={nh}}
\newglossaryentry{d}{
    name={\ensuremath{d}},
    description={design variable},
    value_range={$\mathbb{R}^{\gls{nd}}$},
    sort={d}}
\newglossaryentry{D_set}{
    name={\ensuremath{\mathcal{D}}},
    description={set of~\glsdesc{d}},
    value_range={$\mathbb{R}^{\gls{nd}}$},
    sort={D}}
\newglossaryentry{nd}{
    name={\ensuremath{n_{\gls{d}}}},
    description={dimension of~\glsdesc{d}},
    value_range={$\mathbb{I}^{+}$},
    sort={nd}}
\newglossaryentry{y}{
    name={\ensuremath{y}},
    description={auxiliary variable},
    value_range={$\mathbb{R}^{\gls{ny}}$},
    sort={y}}
\newglossaryentry{ny}{
    name={\ensuremath{n_{\gls{y}}}},
    description={dimension of~\glsdesc{y}},
    value_range={$\mathbb{I}^{+}$},
    sort={ny}}
\newglossaryentry{r}{
    name={\ensuremath{r}},
    description={local response function},
    value_range={$\mathbb{R}^{\gls{nr}}$},
    sort={r}}
\newglossaryentry{nr}{
    name={\ensuremath{n_{\gls{r}}}},
    description={dimension of~\glsdesc{r}},
    value_range={$\mathbb{I}^{+}$},
    sort={nr}}
\newglossaryentry{vf}{
    name={\ensuremath{v_{f}}},
    description={local objective function},
    value_range={$\mathbb{R}$},
    sort={vf}}
\newglossaryentry{vleq}{
    name={\ensuremath{v^{\leq}}},
    description={local inequality constraint function},
    value_range={$\mathbb{R}^{\gls{nvleq}}$},
    sort={vleq}}
\newglossaryentry{vleq_active_set}{
    name={\ensuremath{\mathcal{A}^{\leq}}},
    description={set of active \glsdesc{vleq}},
    value_range={$...$},
    sort={A}}
\newglossaryentry{nvleq}{
    name={\ensuremath{n_{\gls{vleq}}}},
    description={dimension of~\glsdesc{vleq}},
    value_range={$\mathbb{I}^{+}$},
    sort={nvleq}}
\newglossaryentry{veq}{
    name={\ensuremath{v^{=}}},
    description={local equality constraint function},
    value_range={$\mathbb{R}^{\gls{nveq}}$},
    sort={veq}}
\newglossaryentry{nveq}{
    name={\ensuremath{n_{\gls{veq}}}},
    description={dimension of~\glsdesc{veq}},
    value_range={$\mathbb{I}^{+}$},
    sort={nveq}}
\newglossaryentry{c}{
    name={\ensuremath{c}},
    description={coupling constraint function},
    value_range={$\mathbb{R}^{\gls{nc}}$},
    sort={c}}
\newglossaryentry{cc}{
    name={\ensuremath{c_{c}}},
    description={consensus constraint function},
    value_range={$\mathbb{R}^{\gls{ny}}$},
    sort={cc}}
\newglossaryentry{nc}{
    name={\ensuremath{n_{\gls{c}}}},
    description={dimension of~\glsdesc{c}},
    value_range={$\mathbb{I}^{+}$},
    sort={nc}}
\newglossaryentry{N}{
    name={\ensuremath{N}},
    description={set of indices of neighboring subystems},
    value_range={$\mathbb{I}_{0}^{+}$},
    sort={N}}
\newglossaryentry{M}{
    name={\ensuremath{M}},
    description={set of indices of subsystems},
    value_range={$\mathbb{I}_{0}^{+}$},
    sort={M}}
\newglossaryentry{m}{
    name={\ensuremath{m}},
    description={indice of last subsystem},
    value_range={\gls{M}},
    sort={m}}
\newglossaryentry{H}{
    name={\ensuremath{H}},
    description={mapping operator},
    value_range={$\mathbb{R}^{\gls{nh}}$},
    sort={H}}
\newglossaryentry{L}{
    name={\ensuremath{L}},
    description={(augmented) Lagrange function},
    value_range={$\mathbb{R}$},
    sort={L}}
\newglossaryentry{L_c}{
    name={\ensuremath{L_c}},
    description={consensus (augmented) Lagrange function},
    value_range={$\mathbb{R}$},
    sort={L_c}}
\newglossaryentry{weight}{
    name={\ensuremath{s}},
    description={penalty parameter},
    value_range={$\mathbb{R}^{\gls{nc}}$},
    sort={s}}
\newglossaryentry{rho}{
    name={\ensuremath{\rho}},
    description={step size parameter},
    value_range={$\mathbb{R}^{\gls{nc}}$},
    sort={rho}}
\newglossaryentry{lambda}{
    name={\ensuremath{\lambda}},
    description={Lagrange multiplier associated with \gls{c} or \gls{cc}},
    value_range={$\mathbb{R}^{\gls{nc}}$ or $\mathbb{R}^{\gls{ny}}$},
    sort={lambda}}
\newglossaryentry{outer-loop_itr}{
    name={\ensuremath{k}},
    description={outer-loop iterator},
    value_range={$\mathbb{I}_{0}^{+}$},
    sort={k}}
\newglossaryentry{inner-loop_itr}{
    name={\ensuremath{l}},
    description={inner-loop iterator},
    value_range={$\mathbb{I}_{0}^{+}$},
    sort={l}}
\newglossaryentry{u}{
    name={\ensuremath{u}},
    description={coupling parameter},
    value_range={---},
    sort={u}}
\newglossaryentry{p}{
    name={\ensuremath{p}},
    description={coordination parameter},
    value_range={---},
    sort={p}}
\newglossaryentry{P}{
    name={\ensuremath{P}},
    description={coordination objective function},
    value_range={$\mathbb{R}$},
    sort={P}}
\newglossaryentry{Q}{
    name={\ensuremath{Q}},
    description={coordination constraint function},
    value_range={$\mathbb{R}^{\gls{nQ}}$},
    sort={Q}}
\newglossaryentry{nQ}{
    name={\ensuremath{n_\gls{Q}}},
    description={dimension of~\glsdesc{Q}},
    value_range={$\mathbb{I}^{+}$},
    sort={nQ}}
\newglossaryentry{beta}{
    name={\ensuremath{\beta}},
    description={hyperparameter for subgradient method},
    value_range={$[1;\infty[$},
    sort={beta}}
\newglossaryentry{gamma}{
    name={\ensuremath{\gamma}},
    description={hyperparameter for subgradient method},
    value_range={$]0;1[$},
    sort={gamma}}
\newglossaryentry{outer-loop_epsilon}{
    name={\ensuremath{\epsilon_{outer}}},
    description={tolerance for outer-loop convergence criterion},
    value_range={$\mathbb{R}^{+}$},
    sort={epsilon_outer}}
\newglossaryentry{inner-loop_epsilon}{
    name={\ensuremath{\epsilon_{inner}}},
    description={tolerance for inner-loop convergence criterion},
    value_range={$\mathbb{R}^{+}$},
    sort={epsilon_inner}}
\newglossaryentry{dual_residual}{
    name={\ensuremath{\hat{c}}},
    description={dual residual},
    value_range={$\mathbb{R}^{\gls{nc}}$},
    sort={c}}
\newglossaryentry{optimization_function}{
    name={\ensuremath{\pi}},
    description={optimization function of a subsystem},
    value_range={\gls{D_set}},
    sort={pi}}
\newglossaryentry{kappa_leq}{
    name={\ensuremath{\kappa^{\leq}}},
    description={Lagrange multiplier associated with \gls{vleq}},
    value_range={$\mathbb{R}^{\gls{nvleq}}$},
    sort={kappa_leq}}
\newglossaryentry{kappa_eq}{
    name={\ensuremath{\kappa^{=}}},
    description={Lagrange multiplier associated with \gls{veq}},
    value_range={$\mathbb{R}^{\gls{nveq}}$},
    sort={kappa_eq}}
\newglossaryentry{tau}{
    name={\ensuremath{\tau}},
    description={slack variable},
    value_range={$\mathbb{R}^{\gls{nc}}$},
    sort={tau}}
\newglossaryentry{alpha_1}{
    name={\ensuremath{\alpha_{1}}},
    description={globalization step size},
    value_range={$\mathbb{R}_{+}$},
    sort={alpha_1}}
\newglossaryentry{alpha_2}{
    name={\ensuremath{\alpha_{2}}},
    description={globalization step size},
    value_range={$\mathbb{R}_{+}$},
    sort={alpha_2}}
\newglossaryentry{alpha_3}{
    name={\ensuremath{\alpha_{3}}},
    description={globalization step size},
    value_range={$\mathbb{R}_{+}$},
    sort={alpha_3}}
\newglossaryentry{Sigma}{
    name={\ensuremath{\Sigma}},
    description={positive definite scaling matrix},
    value_range={$\mathbb{R}^{\gls{nd}}$},
    sort={Sigma}}
\newglossaryentry{nu}{
    name={\ensuremath{\nu}},
    description={penalty parameter},
    value_range={$\mathbb{R}^{+}$},
    sort={v}}
\newglossaryentry{Box}{
    name={\ensuremath{\Box}},
    description={generalized controller optimization variable},
    value_range={$\mathbb{R}^{\gls{nBox}}$},
    sort={Box}}
\newglossaryentry{nBox}{
    name={\ensuremath{n_{\gls{Box}}}},
    description={dimension of~\glsdesc{Box}},
    value_range={$\mathbb{I}^{+}$},
    sort={nBox}}
\newglossaryentry{vd_active_set}{
    name={\ensuremath{\mathcal{A}_{D}}},
    description={set of active bounds of \glsdesc{D_set}},
    value_range={$...$},
    sort={A_D}}
\newglossaryentry{kappa_D}{
    name={\ensuremath{\kappa_{\mathcal{D}}}},
    description={Lagrange multiplier associated with \gls{vD}},
    value_range={$\mathbb{R}^{\gls{nvD}}$},
    sort={kappa_D}}
\newglossaryentry{S}{
    name={\ensuremath{S}},
    description={selector matrix to retrieve $^{i}\gls{x},\:^{i}_{j}\gls{z},\:^{i}_{j}\gls{h}$ from $^{i}\gls{d}$},
    value_range={---},
    sort={S}}
\newglossaryentry{vD}{
    name={\ensuremath{v_{\mathcal{D}}}},
    description={local bound inequality constraint function},
    value_range={$\mathbb{R}^{\gls{nvD}}$},
    sort={vD}}
\newglossaryentry{nvD}{
    name={\ensuremath{n_{\gls{vD}}}},
    description={dimension of~\glsdesc{vD}},
    value_range={$\mathbb{I}^{+}$},
    sort={nvD}}